# Coherent Virtual Absorption in Dielectric Metasurfaces

Kaizad Rustomji[1], Nasim Mohammadi Estakhri[1,2], Nooshin M. Estakhri[1,2,3]*

[1] *Fowler School of Engineering, Chapman University, Orange, California 92866, USA*

[2] *Schmid College of Science and Technology, Chapman University, Orange, California 92866, USA*

[3] *Institute for Quantum Studies, Chapman University, Orange, California 92866, USA*

*nmestakhri@chapman.edu

**Abstract:** Through temporal shaping of the excitation signal, the complex-frequency scattering zeros of a lossless structure can be accessed, enabling a storage-release mechanism referred to as coherent virtual absorption. Practical demonstrations of this mechanism, however, have been limited to simple configurations such as slabs and spheres, where analytical solutions allow accurate prediction of the complex-frequency scattering zeros. Here, we extend this concept into the realm of metasurfaces and demonstrate coherent virtual absorption in realistic and dispersive metasurface configurations. Through a combination of full-wave analysis and rational approximation, we present a practical scheme to identify suitable complex-frequency zeros and achieve coherent virtual absorption successfully. Our approach can be implemented in arbitrary metasurface configurations with any number of ports, providing a robust framework for optimized energy storage, memories, optical sensing, and modulation in practical photonic systems.

# Introduction

Complex-frequency excitations have opened a temporal avenue for controlling light-matter interactions in photonic structures [1–3]. Rather than sculpting the structure, complex-frequency excitations have been used to mimic *effective* virtual gain and loss in simple and otherwise lossless scattering systems, enabling phenomena such as super-resolution [4], critical coupling [5], and extreme scattering [6]. Using scattering matrix (S-matrix) notation [7], complex-frequency excitation signifies operating at the location of a zero (i.e., virtual loss) or pole (i.e., virtual gain) of the eigenvalue of the scattering matrix by excitation with exponentially growing or decaying pulses. For instance, when operating at the frequency of a complex zero of the system, incident wave(s) can be properly tuned such that all outgoing fields destructively interfere, resulting in coherent virtual absorption [8–10]. Under this condition, the system continuously accumulates energy without reflection or transmission, with the total stored energy within the structure increasing over time until the excitation is terminated. This mechanism can enable dynamic energy storage [8,11–13], enhanced molecular sensing [14], and manipulate optical forces [15]

Virtual absorption and gain have been studied in simple geometries where analytical or semi-analytical solutions allow accurate prediction of the complex spectral behavior of the S-matrix. Instances of such structures include planar geometries such as single or multi-layered slabs [3,8,16], spherical objects [17–19], and configurations that can be approximated with a lumped-element model [13,20,21]. However, extension of this concept to arbitrarily shaped geometries is not trivial. In a general inhomogeneous, dispersive, and multimode system, it is typically very hard, if not impossible, to analytically calculate the scattering poles and zeros of the system in the complex frequency space. Moving beyond these ideal and unrealistic cases, it is essential to develop general and reliable techniques to access and understand complex-frequency effects in more practical scenarios [20–24]

In this work, we present the successful realization of coherent virtual absorption effect in arbitrary and realistic dielectric metasurfaces without any approximation or pre-assumptions. We report two distinct scenarios: a semi-randomized metasurface with high-index inclusions embedded within a low-index background (Fig. 1a), and a patterned metasurface with inclusions positioned on a common substrate (Fig. 4a). In both cases, the thickness of the metasurfaces is chosen to be optically large, thus eliminating the possibility of modeling as effective sheet impedance or allowing for transmission-line treatment. The patterned metasurface employs silicon (Si) and silicon dioxide ($SiO_2$) with realistic dispersion, demonstrating a low-loss infrared platform 1.4 $\mu$m − 2.7 $\mu$m ) well-suited for the experimental realization of coherent virtual absorption. By combining time-domain numerical simulations [3,22], analytic continuation techniques [23–25], and spectral filtering [3], we accurately identify scattering zeros of the system. The analysis is followed by a full characterization of the scattering response of the system under temporal excitations mimicking the relevant complex frequencies. Our approach is suitable for multi-port metasurface configurations, helping to bridge the gap between theoretical complex-frequency excitations and their practical implementation.

## Results and Discussion

We start by considering an inhomogeneous metasurface, as shown in Fig.1a, under normal plane wave illumination from both sides. The unit cell dimensions are chosen to ensure that oblique scattering modes remain evanescent over the entire frequency range of interest, creating a 2-port system. In view of complex-frequency analysis, the input waves are assumed to have a complex frequency $\omega = \omega' + i\omega''$, with electric fields polarized along the $y$-axis. Assuming that the system can be fully characterized in the complex-frequency domain, the zeroes of the eigenvalues of the S-matrix indicate the infinite set of complex frequencies associated with complete (virtual) absorption. Under the $e^{-i\omega t}$ time convention in the temporal analysis, the incident waves can then be tailored to transiently mimic the complex excitation and switching off after a certain amount of time. The incident wave(s) take the general form of $E_{inc}(t) = E_0 e^{\omega'' t} e^{-i\omega' t} u(-t) + E_0 e^{-(t/\tau)^2} e^{-i\omega' t} u(t)$, with $\tau$ adjusting the decay rate after the main signal is switched off at $t = 0$ [8,16]. The Gaussian decay for $t \geq 0$ ensures a smooth cutoff after the pulse is switched off. Here we set the switch-off time at $\tau = 1/\omega'$, indicating that the field smoothly decays within two oscillations. With the switching time set at $t = 0$, we also define $t = t_0$ (Fig. 1a) as the time at which the maxima of the incident pulse envelopes reach the metasurface edges located at $x = \pm L/2$. Regardless of the internal composition of the metasurface, the structure is always symmetrically positioned within this spatial range, and at this exact time, the Gaussian portions of the incident waves start to interact with the metasurface, eliminating the coherent virtual absorption condition. After this point in time, scattering from the metasurface is expected.

Given the complex nature of wave-matter interaction in an arbitrarily shaped metasurface with an optically large thickness, the scattering behavior of the system cannot be analytically modeled. As such, to calculate the eigenvalues of the S-matrix in the complex frequency plane, the scattering response of the system is first numerically analyzed in the real frequency domain. Indeed, and in most practical situations, we need to rely on numerical or experimental data available over a limited and often noisy range of real frequencies for such extensions [23,26]. Targeting a complex zero of the scattering matrix of the system, we start by numerically modeling the structure using a time domain analysis technique [22]. The choice of a time domain approach allows us to also monitor the evolution of the wave interaction with the metasurface in time, providing valuable information on the onset of scattering and the impact of switching cut-off time [3]. The calculated 2-port S-parameters [7] are analytically continued into the complex-frequency plane using the Adaptive Antoulas–Anderson (AAA) algorithm [27–29] with a maximum of 100 iterations and a relative tolerance of 0.1. The AAA algorithm allows us to build rational approximations directly from discrete data, reliably capturing zeros and poles with strong numerical stability. The reference planes are chosen at the edges of the metastructures, i.e., $x = \pm L/2$, to reduce the potential occurrence of spurious zeros and poles in our problem [29]. In all cases, the extrapolated complex frequency scattering parameters are superimposed on the original data to ensure proper accuracy on the real axis. This is shown in Fig. 1b, comparing S-parameters across real frequency corresponding to $5 < \text{Re}(kL) < 8.6$, which lies within the bandwidth of the incident Gaussian pulse. The results indicate very high accuracy, implying precise prediction of the complex-frequency behavior of the system.

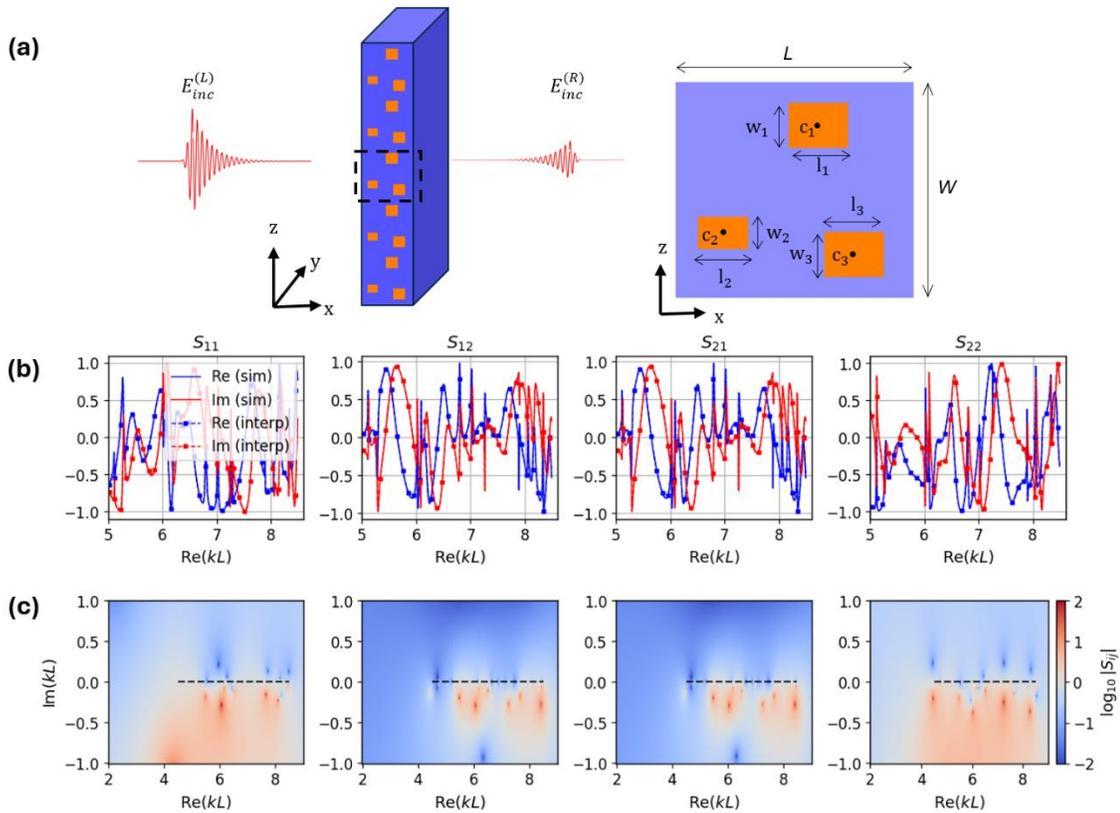

**Figure 1** (a) Schematic illustration of the randomized dielectric metasurface illuminated by two counterpropagating pulses. The background refractive index is set at $n = 2$ (blue region) with inclusions of index $n = 5$ (orange regions). The width of the unit cell, i.e., periodicity in the z direction, is set to $W = 4.8/k_0$ and the thickness of the metasurface is $L = 6.93/k_0$ with $k_0 = 2\pi/\lambda_0, \lambda_0 = 0.6$ μm. The centers of inclusions are placed at $c_1 = (0.1L, 0.3W), c_2 = (-0.3L, -0.2W), c_3 = (0.25L, -0.3W)$ with $w_1 = 0.21, w_2 = 0.151, w_3 = 0.21$ and $l_1 = 0.251L, l_2 = 0.21L, l_3 = 0.251L$, as illustrated on the right panel. (b) Numerically obtained S-parameters vs normalized frequency. In each panel, red and blue colors indicate the real and imaginary parts of scattering elements (solid lines). Discrete dots indicate the values of S-parameters on the real frequency axis retrieved after applying the AAA algorithm. (c) Full picture of the interpolated S-parameters obtained from the AAA algorithm shown in the normalized complex-frequency plane. The black dashed line denotes the sampling line on the real frequency axis used to extrapolate the data. Red and blue hotspots indicate scattering the estimated poles and zeros, respectively.

Relying on the extrapolation accuracy, the full complex-frequency response of the metastructure is shown in Fig. 1c, with the black dashed line indicating the real-frequency input data range. Multiple S-parameter complex zeros and poles are identifiable with some spurious points near the edges of the provided frequency range and further from the real axis. We note that the eigenvalues far from the real axis are not of interest as the corresponding excitation will exhibit very rapid attenuation before any meaningful interaction with the metastructure. The two eigenvalues of the S-matrix are then calculated at each frequency, and the zeroes and poles of the scattering response are numerically estimated by locating the local minima and maxima. This data is illustrated in Fig. 2, indicating the presence of multiple pairs of zeros and poles in the complex frequency plane. Numerical noises, incomplete spectral sampling, and limited frequency resolution can indeed produce spurious features. Conjugate pairs of poles and zeros are generally expected for a lossless system with all the zeros placed in the upper half-plane. While potential numerical artifacts such as Froissart doublets [28] can occur, increasing the accuracy of the input data and proper algorithm setup minimizes their impact.

To demonstrate coherent virtual absorption, we select one of the identified complex-frequency zeros of the eigenvalue set in Fig. 2, marked by a green loop. The zero is positioned at $kL = 6.04 + 0.31i$, setting a ratio of $\omega'/\omega'' \approx 19.48$. This choice is intentional and is to achieve a reasonable complex-frequency response without excessive attenuation, as described above. Given the inherent asymmetry of the system (looking from left to right vs right to left), we also expect an asymmetric associated eigenvector, which is found to satisfy $E_{inc}^{(R)}/E_{inc}^{(L)} = 10.48\angle 165.9°$. Interestingly, we note that as the position and size of the inclusions are chosen randomly, the eigenvector does not significantly deviate from a fully symmetric slab, and the ratio of opposite waves can get closer to

±1 by increasing randomness, with interesting implications for achieving coherent virtual absorption in wavelength-scale randomized structures [30–32] and periodically arranged random configurations [33].

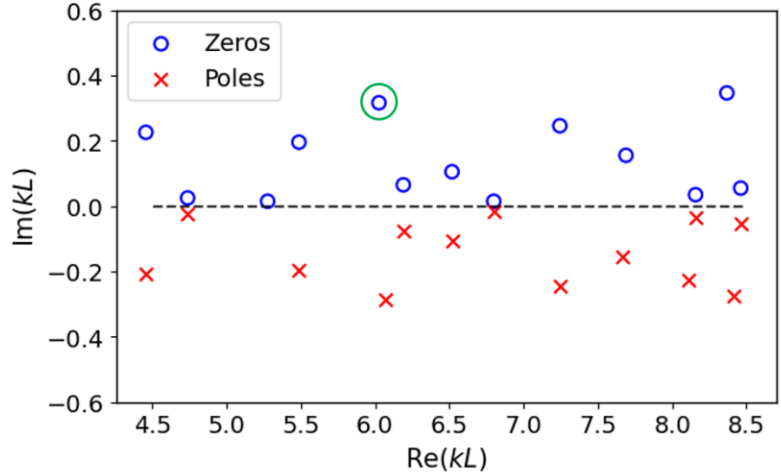

**Figure 2** Distribution of the zeros and poles of the eigenvalues of the S-parameters in the complex frequency plane. The green loop marks the chosen complex-frequency zero located at $kL = 6.04 + 0.31i$ and the black dashed line indicates the input range of the real frequency.

Excitation of the system at this complex frequency and with the associated eigen vector is expected to entirely suppress scattering from the randomized slab, up to the point in time that the incident wave is switched off. Accordingly, we define two reference time parameters $t_0$ and $t_{monitor}$. $t_0$ refers to the moment when the maxima of the incident pulse envelopes (initiated from the ports placed well outside of the near field region) reach the edges of the slab. At this time, the non-complex-frequency portion of the incident pulses start to interact with the randomized metasurface. $t_{monitor}$, on the other hand, refers to the time that the maxima of the incident pulse(s) reach the location of observation monitor(s) placed outside of the slab after reflecting from the metasurface. The monitors are placed at $x_{\text{monitor}} = \pm L/2 \pm 4/k'$, sufficiently far from the randomized metasurface to avoid near-field contributions. Results presented in Figs. 3 and 5 are evaluated until $t_{monitor}$, corresponding to the arrival of the leading edge of the pulse reflected from the slab.

Fig. 3 illustrates the successful pinpointing of the coherent virtual absorption condition. Fixing the complex frequency of the excitation waves, we gradually change the phase offset between the excitation fields on the left and right sides while keeping the amplitude ratio fixed at 1.048. Using the fields directly available from the FDTD simulations, the total scattered energy from each side

$\int_0^{t_{monitor}} \text{Re}\,(E_y^{\text{scat}}(x_{monitor}, t))\text{Re}(H_z^{\text{scat}}(x_{monitor}, t))\,dt$ is estimated, as shown by solid lines in Fig. 3a. As expected, at the phase offset of approximately 167° the scattered energy is minimized from both sides. Interestingly, and despite the semi-randomized configuration of the elements and relative excitation amplitude of approximately one from the two sides, breaking the coherent virtual absorption condition does not symmetrically affect the scattering from different sides. Indeed, we note that this variation is rooted in the configuration not being fully randomized and consisting of three elements per period; however, the strong manifestation of this asymmetry in the phase dependency is intriguing. Fig. 3a also illustrates the same results now calculated using a spectral filtering approach, utilizing the space-time Fourier transform technique to isolate the scattered energy on each side [3]. Similarly, Fig. 3b illustrates the trajectory of the scattered energy versus the complex frequency of the excitation wave, where the imaginary part of the complex excitation frequency $\omega''$ is gradually modified across the coherent virtual absorption point (upward and downward). The results confirm the pinpointing of the coherent virtual absorption condition, indicated with the vertical dashed lines.

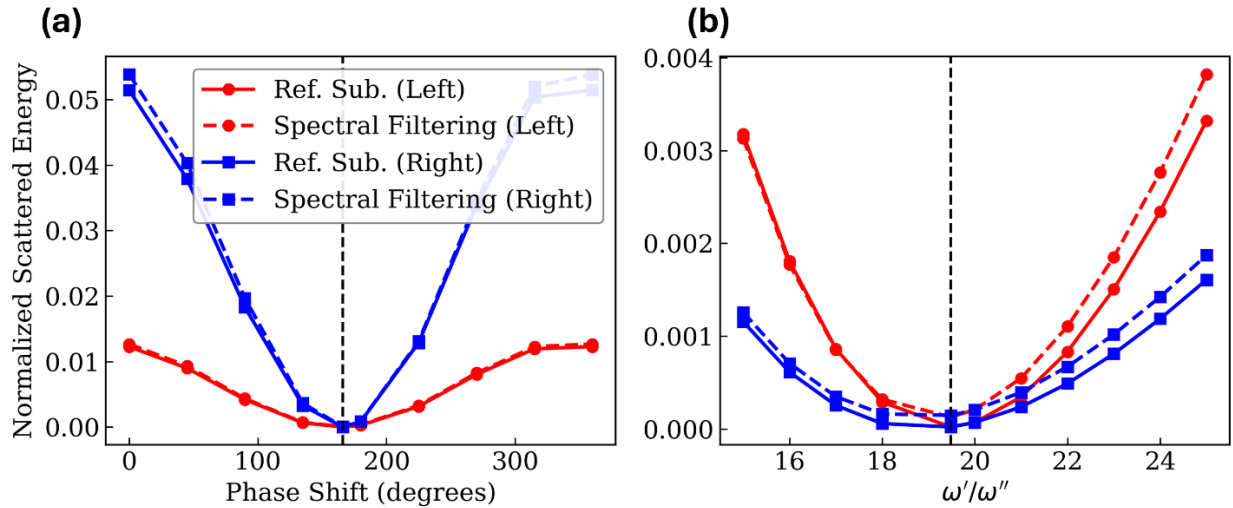

Figure 3. Normalized scattered energy versus (a) the phase shift between two incident pulses and (b) variations in the excitation complex frequency. Red and blue colors indicate the normalized scattering from the left and right sides of the randomized metasurface, respectively. Results are obtained using the fields values directly obtained from FDTD simulations (solid lines) as well as the spectral domain filtering method (dashed line) [3]. The vertical lines in each panel indicate the expected coherent virtual absorption condition.

With the successful demonstration of coherent virtual absorption in the randomized slab, we now consider a more realistic scenario with a patterned dispersive metasurface. The configuration of the metasurface is illustrated in Fig. 4a, representing an electromagnetically thick structure operating in the near infrared regime. As mentioned above, the thickness is intentionally chosen such that the metasurface cannot be modeled as an ultrathin load, consequently, and in conjunction with the surface inhomogeneity, providing a general and representative platform for studying coherent virtual absorption beyond ideal configurations [3,8,16]. The periodicity of the metasurface is set to 1.3 $\mu$m, aiming for the wavelength operation range of $\approx 1.4\ \mu m - 2.7\ \mu m$ where both Si and SiO2 are low loss and exhibit moderate dispersion [34]. The low-loss regime is intentionally chosen to ensure that any observed absorption arises from coherent virtual absorption rather than material losses. The configuration is well-suited for experimental realization of coherent virtual absorption with the metasurface geometry compatible with standard semiconductor fabrication techniques, such as electron-beam lithography [35] and focused ion beam milling [36,37].

The computational steps for this metasurface are consistent with the previous structure as explained before and the extracted zeroes and poles are accordingly shown in Fig. 4b. Several zero-pole pairs are identified in the frequency range of interest, and we also notice occurrence of double poles or spurious singularities, particularly beyond the range of the input real frequency. This is expected and we envision a higher accuracy through increasing the sampling rate on the real axis and iterative variations in the algorithm setup to reduce or remove these artifacts. Coherent virtual absorption is then demonstrated for the identified complex-frequency zero, marked in Fig. 4b by a green loop. This zero is positioned at $kL \approx 9.24 + 06i$ with a complex frequency of $\omega = \omega' + i\omega'' \approx (1.018 + i0.066) \times 10^{15}$ Hz and a ratio of $\omega'/\omega'' = 15.419$. We are once more cautious to avoid excessive attenuation (i.e., avoiding small $\omega'/\omega''$ ratios) yet still choosing zeros far enough from the real frequency axis, preventing long computation runtimes as we observe the coherent virtual absorption effect.

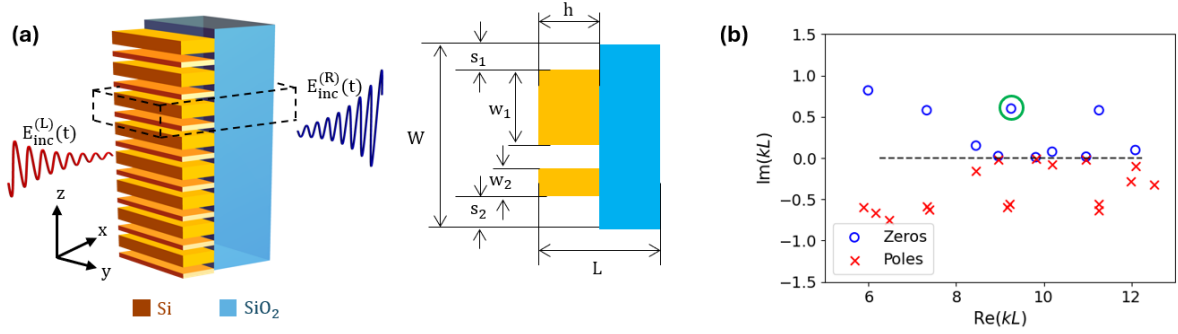

**Figure 4** (a) Schematic illustration of the dispersive dielectric metasurface illuminated by two counterpropagating pulses from left and right. Metasurface consists of silicon elements (orange) positioned on a silicon dioxide substrate (blue). Width of the unit cell, i.e., periodicity in the z direction, is set to $W = 1.3\ \mu m$ and the total thickness of the metasurface is $L = 2.72\ \mu m$. Other parameters are set at $h = 0.43\ \mu m, s_1 = 0.14\ \mu m, s_2 = 0.14\ \mu m, w_1 = 0.57\ \mu m, w_2 = 0.22\ \mu m$, as illustrated in the right panel. (b) Distribution of the zeros and poles of the eigenvalues of the S-parameters in the complex frequency plane for the configuration shown in (a). The green loop marks the chosen complex-frequency zero at $kL = 9.24 + 0.6i$ and the black dashed line indicates the input range of the real frequency.

Compared to the previous scenario of the randomized slab, a more apparent structural asymmetry is present in the dielectric metasurface in Fig. 4a, and consequently, we expect more pronounced asymmetric behavior from the associated eigenvectors. This is indeed true, and the eigenvector is found to satisfy $E_{inc}^{(R)}/E_{inc}^{(L)} = 1.95\angle 190.87°$ with the wave incident from the substrate side to be almost twice in amplitude and opposite in sign compared to its coherent pair incident from the left side on the metasurface. The time evolution of the electric field in the vicinity of the metasurface is shown in Fig. 5, plotted for various instances before and after the maxima of the incident beams reach the edges of the metasurface. We note that in an ideal scenario, an abrupt discontinuation of the complex frequency excitation will promptly initiate the scattering. Gradual termination of the field can slightly delay the onset of scattering, as extensively studied in [3] and observed here.

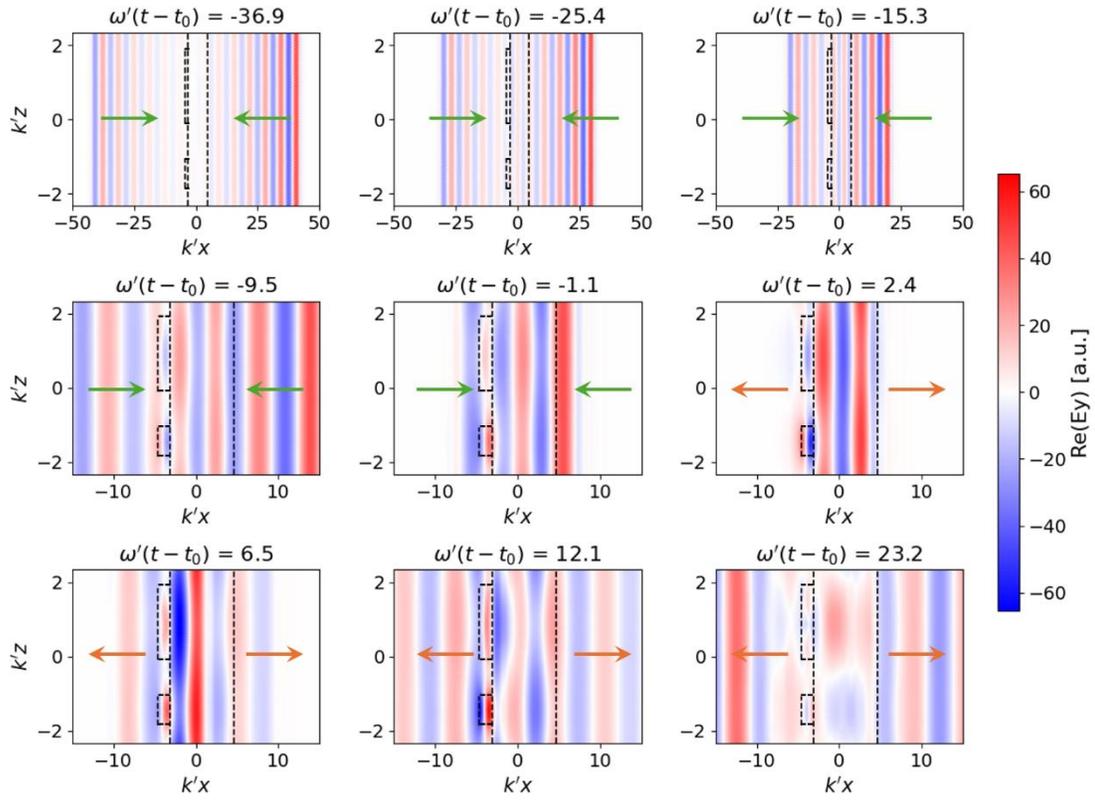

Figure 5: Temporal volution of electric field in the vicinity of the metasurface. The arrows are visual guides to better demonstrate the propagation and scattering.

Finally, Fig. 6 demonstrates the observation of coherent virtual absorption conditions, where we investigate the gradual removal of the coherent absorption condition through variations in the phase offset of the incident beams in Fig. 6a and $\omega'/\omega''$ in Fig. 6b, confirming the findings.

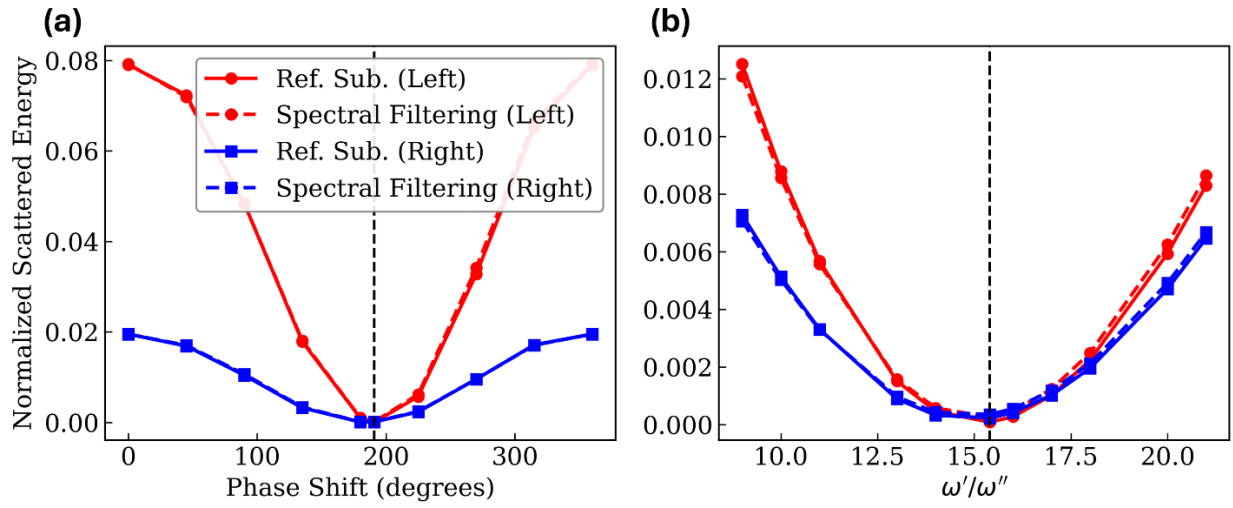

Figure 5. Normalized scattered energy versus (a) the phase shift between two incident pulses and (b) variations in the excitation complex frequency. The vertical lines in each panel indicate the expected coherent virtual absorption condition.

## Methods

The Flexcompute Tidy3D simulation software is used to perform all FDTD simulations with complex-frequency excitation beams [22]. Separate simulations were also carried out for S-parameter extraction where the two counter-propagating plane-wave sources were placed symmetrically at $\pm 25/k_0$ with $k_0 = 2\pi/\lambda_0$ where $\lambda_0$ indicates the central excitation wavelength. The two incident waves normally illuminate the centrally located metasurface, with the electric field polarized along the $y$-axis and waves propagation along the $\pm x$-direction. A uniform mesh size of $\lambda_0/80$ is used in all simulation and the bandwidth of excitation signals (for S-parameter calculations) is set to $f_0/7$. Central wavelengths are $\lambda_0 = 0.6\ \mu m$ for randomized slab simulations, i.e., Figs. 1-3, and $\lambda_0 = 1.8\ \mu m$ for the infrared metasurface simulations, i.e., Figs 4-6. Bloch periodic boundary conditions are applied in the $z$-direction, and the domain is truncated with perfectly matched layers in the $x$-direction. For complex frequency excitations, the sources are placed farther from the structure and at $\pm 280/k'$. Simulation time step and maximum simulation runtime are set to $\Delta t = 1.74 \times 10^{-5}$ and 12 ps for the randomized metasurface and $\Delta t = 5.21 \times 10^{-5}$ and 5 ps for the infrared metasurface, with simulations terminating if the instantaneous integrated E-field intensity drops below $10^{-5}$, before reaching the maximum runtime. Simulations corresponding to the demonstration of coherent virtual absorption, i.e., complex-frequency excitation, are performed based on the extracted complex-frequency zeros. For the randomized slab $\lambda' = 0.6\ \mu m$ and for the dispersive infrared metasurface $\lambda' = 1.851\ \mu m$.

Each individual S-parameter $S_{ij}(\omega)(i,j \in \{1,2\})$ is obtained from FDTD simulations at real-valued frequencies and then interpolated into the complex frequency plane using the AAA algorithm [27,28], with a relative tolerance of 0.1 and a maximum of 100 iterations. The resulting complex-frequency interpolated functions $\tilde{S}_{11}, \tilde{S}_{12}, \tilde{S}_{21}$, and $\tilde{S}_{22}$ are re-assembled to obtain the extended S-matrix in the complex plane, whose eigenvalues $\lambda_1$ and $\lambda_2$ are computed. In all cases, the eigenvalues are deliberately labeled such that $|\lambda_1| < |\lambda_2|$. In this convention, $\lambda_1$ contains the zeros of the eigenvalues of the S-matrix, while $\lambda_2$ contains the corresponding poles. A local numerical search in the complex frequency plane is performed to locate minima of $|\lambda_1|$ and maxima of $|\lambda_2|$, determining the locations of the S-matrix zeros and poles, respectively. Summarized graphs are presented in Figs. 2 and 4d.